\documentclass[showpacs,aps,graphicx,twocolumn]{revtex4}

\usepackage{graphicx,CJK}

\begin{document}

\title{Phononic entanglement concentration via optomechanical interactions}

\author{Shan-Shan Chen, Hao Zhang,\footnote{zhanghaoguoqing@163.com} Qing Ai, and Guo-Jian Yang\footnote{yanggj@bnu.edu.cn}}

\address{Department of Physics, Applied Optics Beijing Area Major Laboratory, Beijing Normal University, Beijing 100875, China}

\date{\today}

\begin{abstract}
Low dissipation, tunable coupling to other quantum systems, and unique features of phonons in the aspects of propagation, detection and others suggest the applications of quantized mechanical resonators in phonon-based quantum information processing (QIP) in a way different from their photonic counterpart. In this paper, we propose the first protocol of entanglement concentration for nonlocal phonons from quantized mechanical vibration. We combine the optomechanical cross-Kerr interaction with the Mach-Zehnder interferometer and, by means of twice optomechanical interactions and the photon analysis with respect to the output of the interferometer, achieve ideal entanglement concentration about less-entangled nonlocal phonon Bell and Greenberger-Horne-Zeilinger states. Our protocol is useful for preserving the entangled phonons for the use of high quality phonon-based QIP in future.
\end{abstract}

\maketitle

\section{Introduction}\label{sec1}
Quantum entanglement is a quantum resource indispensable in the areas, such as quantum key distribution \cite{EkertPRL1991,BennetPRL1992,DengPRA2008}, quantum teleportation \cite{BennetPRL1993}, quantum secure direct communication \cite{LongPRA2002,DengPRA2003,WangPRA2005,ZhangPRL2017} and quantum dense coding \cite{BennettPRL1992,LiuPRA2002}. In quantum communications, an entangled state is usually used for building a quantum channel between remote parties. However, the channel noise will induce decoherence and degrade the entanglement between the quantum systems. Entanglement concentration \cite{BennettPRA1996,YamamotoPRA2001,ZhaoPRA2001,ShengPRA20081,LiPRA2015,BoteroPRA2018,BaoPRA2013,ShengPRA20121,
DengPRA20121,WangPRA2012,ZhangPRA2016,Zhangaop2018,lichunyanoe2019} is an operation which converts a partially entangled state to a more or maximally entangled state. Since the first entanglement concentration protocol (ECP), the well known Schmidt projection, was proposed by Bennett \emph{et al} \cite{BennettPRA1996}, its realization on the various physical systems has been reported. The examples include ECPs with linear optical elements in the principle of Schmidt projection \cite{YamamotoPRA2001,ZhaoPRA2001},  nonlocal-photon ECP with linear \cite{BaoPRA2013} or nonlinear optical elements \cite{ShengPRA20121,DengPRA20121}, and ECP based on electron-spin systems \cite{WangPRA2012} or on circuit quantum electrodynamics \cite{ZhangPRA2016,Zhangaop2018}, etc.

Mechanical resonators are known as important platforms on which one can generate different quantum effects \cite{XuhuNature2019,XuhuNature2016,FlayacPRL2014,WoolleyPRA2014,LiNJP2015,Ockeloen2018,RiedingerNature2018} and realize quantum information processing (QIP) \cite{MAspelmeyerRMP2014,LudwigPRL2012}. As the information carriers of quantized mechanical vibration modes, phonons can be confined in mechanical resonators and propagate in phononic waveguides \cite{YuAPL2014,TsaturyanNN2017, Patelprl2018,MeenehanPRX2015}. Compared to photons, the phonons' speed is much slow, thus they are more suitable for storaging and transferring quantum information between quantum nodes over a short distance. With the development of the technique of phononic crystals, which can effectively enhances the mechanical Q \cite{YuAPL2014,TsaturyanNN2017}, and due to the distinct advantages with low dissipation and tunable coupling to other systems, mechanical resonators and their complex, such as optomechanical systems \cite{MAspelmeyerRMP2014,LudwigPRL2012,StannigelPRL2012,WangPRL2012,TianPRL2012,ChanNature2011,YCLiuPRL,FCLeiOE,HZHANGOE2019} and hybrid solid phonon-spin systems \cite{MacQuarrielPRL2013,OvartchaiyapongNC2014,GolterPRL2016,GolterPRX2016,MacQuarriegNC2017}, have been applied to storage, transducers and sensors in QIP \cite{GustafssonScience2014,MJAPRX2015,SafaviNaeiniNJP2011,OkamotoNPhys2013}, and to build scaling-on-chip phononic quantum networks \cite{MarkPRX2018,LemondePRL2018}.

Phonon entanglement is generated with quantized mechanical vibration modes and, on this subject, there are already some important works \cite{FlayacPRL2014,LiNJP2015,Ockeloen2018,RiedingerNature2018}. In this paper we focus on how to preserve the phonon entanglement based on the concept of entanglement concentration. The most existing ECPs were designed for photons, which work with linear or nonlinear optical elements and photon-detectable instruments. Because there are not phononic linear elements, phonons cannot be operated as photons can, and the photonic ECPs do not apply to the phonons. In the present investigation, we propose the first ECP for entangled nonlocal phonons. In this protocol, phonons are operated indirectly by controlling photons via optomechanical interaction, so that the original partially entangled two-phonon state is converted to the Bell state or, similarly, a partially entangled three-phonon state to the Greenberger-Horne-Zeilinger (GHZ) state. In details, 4 steps need taking in our ECP. In step 1, using the combination of an cross-Kerr optomechanical system with a Mach-Zehnder interferometer, we obtain the maximally entangled phonon state with a certain successful probability, where the dimension of this phonon state is two times as large as that of the original state to be concentrated for entanglement. In step 2, using the second optomechanical interaction to drive the anti-Stokes transition, we map the state of the whole system onto a phonon-photon state. Then, after all the users, including Alice, Bob and so on, make a Hadamard gate operation on their own photons in step 3 and detect photons with photondetectors as well as share the detection results in step 4, we complete the entanglement concentration about that less-entangled multi-phonon state.

This article is organized as follows: In Sec.~\ref{sec2}, we give a description of optomechanical measurement and show how a partial-entangled nonlocal two-phonon state becomes a Bell state by applying our ECP. In Sec.~\ref{sec3}, we extend our ECP to the multiuser GHZ state and give two simple remarks about the feasibility of our protocol. A summary is given in Sec.~\ref{sec4}.


\begin{figure}[!ht]
\begin{center}
\includegraphics[width=8.0cm,angle=0]{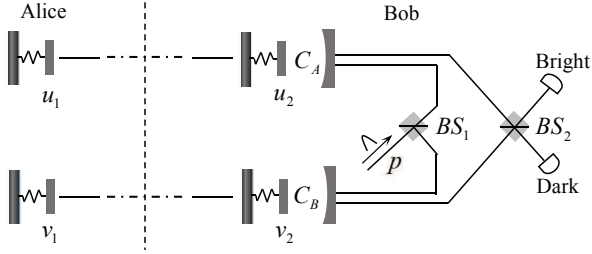}
\caption{ Schematic diagram of step 1 used in the ECP. $u_{j}$ ($v_{j}$) ($j=1,2$) are 4 mechanical oscillators, among which $u_2$ ($v_2$) acts as the end-mirror of the Fabry-Perot cavity $C_{A}$ ($C_{B}$). $BS_{1}$ is a beam splitter through which a photon $p$ is sent into $C_{A}$ and $C_{B}$ at equal probability. Beam splitter $BS_{2}$ together with $BS_{1}$ and two cavities constitute a Mach-Zehnder interferometer used for orthogonal postselection. All elements on the left of vertical dashed line are used by Alice, while those on the right by Bob.}\label{fig1}
\end{center}
\end{figure}

\section{Entanglement concentration for partial-entangled nonlocal phonons}\label{sec2}
We assume that the partial-entangled nonlocal phonons under the consideration of entanglement concentration are in a state shared by two remote mechanical resonators $b_{1}$ and $b_{2}$, and fulfils
\begin{eqnarray}      \label{eqH1}
|\psi\rangle_{b_{1}b_{2}}=\alpha|10\rangle_{b_{1}b_{2}}+\beta|01\rangle_{b_{1}b_{2}},
\end{eqnarray}
with $|\alpha|^{2}+|\beta|^{2}=1$, where $|i_1i_2\rangle_{b_{j_{1}}b_{j_{2}}}$ ($i_1,i_2=0,1$ and $j_1,j_2=1,2$) is the phonon number state with $i_1$ phonons in the resonator mode $b_{j_1}$ and $i_2$  phonons in the resonator mode $b_{j_2}$. For simplicity, we have used the abbreviation $|i_1\rangle_{b_{j_1}}|i_2\rangle_{b_{j_2}}...|i_k\rangle_{b_{j_k}}=|i_1i_2...i_k\rangle_{b_{j_1}b_{j_2}...b_{j_k}}$ to compact the expression of a multimode state of resonators. This is also applied to the expression of a multimode state of optical cavities, but we always write a phonon state separately with a photon state. State Eq.(\ref{eqH1}) results from the decoherence of the ideal Bell state $|\psi\rangle_{b_{1}b_{2}}=\frac{1}{\sqrt{2}}(|10\rangle_{b_{1}b_{2}}+|01\rangle_{b_{1}b_{2}})$ (see Appendix for the generation of this state) due to the fact that the mechanical resonators $b_1$ and $b_2$ operating in dissipative environment or interacting with other quantum systems. The target state after concentration is that with
$\alpha=\beta=\frac{1}{\sqrt{2}}$. Four steps are designed in our phononic ECP. In step 1, as shown by Fig.~\ref{fig1}, we introduce an auxiliary phonon state which is exactly the same as the state, Eq.(\ref{eqH1}), to be concentrated. The phonons in these two states come from two pairs of resonator modes $u_1v_1$ and $u_2v_2$ held respectively by Alice and Bob, and are described by
\begin{eqnarray}      \label{eqinitial}
&&|\psi\rangle_{u_{1}u_{2}}=\alpha|10\rangle_{u_{1}u_{2}}+\beta|01\rangle_{u_{1}u_{2}},\nonumber\\
&&|\psi\rangle_{v_{1}v_{2}}=\alpha|10\rangle_{v_{1}v_{2}}+\beta|01\rangle_{v_{1}v_{2}}.
\end{eqnarray}
It is now needed for Bob to obtain from Eq.(\ref{eqinitial}) a maximally entangled 4-phonon state by performing single photon postselection via the optomechanical interaction. To this end, considering an optomechanical system with a cross-Kerr interaction between the mechanical resonator and the optical cavity which, in the rotating frame, is describe by the Hamiltonian \cite{ChenOE2017,YinOL2018,Liaoarxiv2018}
\begin{eqnarray}      \label{eqH03}
\hat{H}=\Delta \hat{c}^{\dag}\hat{c} + \omega_{m}\hat{b}^{\dag}\hat{b} - g\hat{c}^{\dag}\hat{c} \hat{b}^{\dag}\hat{b},
\end{eqnarray}
where $\hat{c}^{\dag}(\hat{c})$ and $\hat{b}^{\dag}(\hat{b})$ are the creation (annihilation) operators for the cavity and the mechanical resonator, respectively. $\Delta$ is the effective mechanically modulating detuning of the cavity,  $\omega_{m}$ is the mechanical frequency of the resonator and $g$ denotes the effective coupling between the mechanical resonator and the cavity. Under the action of Hamiltonian Eq.(\ref{eqH03}), a Fock state of a phonon-photon system evolutes with the phase accumulation in the way

\begin{eqnarray}      \label{eqH04}
&&|0\rangle _{c}|0\rangle _{_{b}}\longrightarrow |0\rangle _{c}|0\rangle _{_{{b}}},\nonumber\\
&&|0\rangle _{c}|1\rangle _{_{{b}}}\longrightarrow e^{-i\theta_{01}}|0\rangle _{c}|1\rangle _{_{{b}}},\nonumber\\
&&|1\rangle _{c}|0\rangle _{_{{b}}}\longrightarrow e^{-i\theta_{10}}|1\rangle _{c}|0\rangle _{_{{b}}},\nonumber\\
&&|1\rangle _{c}|1\rangle _{_{{b}}}\longrightarrow e^{-i\theta_{11}}|1\rangle _{c}|1\rangle _{_{{b}}},
\end{eqnarray}
where the $|0\rangle _{c}$ ($|1\rangle _{c}$) and $|0\rangle _{_b}$ ($|1\rangle _{_b}$) represent the ground (first excited) state of cavity mode $c$ and mechanical resonator mode $b$, respectively. Here, we have introduced the phases $\theta_{01}=\omega_{m}t$, $\theta_{10}=\Delta t$ and $\theta_{11}=(\omega_{m}+\Delta-g)t$.

By applying the operation Eq.(\ref{eqH04}) to state Eq.(\ref{eqinitial}), that is, as shown in Fig.~\ref{fig1}, letting a photon entering the cavity $A$ or $B$, the state of the whole phonon-photon system will evolve into
\begin{widetext}
\begin{eqnarray}      \label{eqH}
|\psi(t)\rangle&=&\frac{1}{\sqrt{2}}(|01\rangle_{AB}+|01\rangle_{AB})\otimes(\alpha|10\rangle +\beta|01\rangle)_{_{u_{1}u_{_{2}}}}\!\!\!\otimes(\alpha|10\rangle+\beta|01\rangle)_{_{v_{1}v_{2}}} ,\nonumber\\
\longrightarrow&&\frac{1}{\sqrt{2}}[|10\rangle_{AB}\otimes(\alpha e^{-i\theta_{10}}|10\rangle+\beta e^{-i\theta_{11}}|01\rangle)_{_{u_{1}u_{_{2}}}}\otimes(\alpha |10\rangle+\beta e^{-i\theta_{01}}|01\rangle)_{_{v_{1}v_{_{2}}}}\nonumber\\
&&+|01\rangle_{AB}\otimes(\alpha |10\rangle +\beta e^{-i\theta_{01}}|01\rangle)_{_{u_{1}u_{_{2}}}}
\otimes(\alpha e^{-i\theta_{10}}|10\rangle+\beta e^{-i\theta_{11}}|01\rangle)]_{_{v_{1}v_{_{2}}}},
\end{eqnarray}
\end{widetext}
where $|i_1i_2\rangle_{AB}$ ($i_1,i_2=0,1$) means the photon state with $i_1$ ($i_2$) photons in cavity $A$ ($B$). If an photon is detected at the dark port after the second beam splitter $BS_{2}$, an orthogonal postselection is performed with the single-photon state $|\psi_{f}\rangle =\frac{1}{\sqrt{2}}(|10\rangle_{AB}-|01\rangle_{AB})$. Then, the final state for the mechanical resonators becomes
\begin{eqnarray}      \label{eqH}
|\psi\rangle_{1}&=&\frac{1}{2}\Big[(\alpha e^{-i\theta_{10}}|10\rangle+\beta e^{-i\theta_{11}}|01\rangle)_{_{u_{1}u_{_{2}}}}\nonumber\\
&&\otimes(\alpha |10\rangle+\beta e^{-i\theta_{01}}|01\rangle)_{_{v_{1}v_{_{2}}}}\nonumber\\
&&-(\alpha |10\rangle+\beta e^{-i\theta_{01}}|01\rangle)_{_{u_{1}u_{_{2}}}}\nonumber\\
&&\otimes(\alpha e^{-i\theta_{10}}|10\rangle+\beta e^{-i\theta_{11}}|01\rangle)\Big]_{_{v_{1}v_{_{2}}}}\nonumber\\
&=&\frac{1}{2}\alpha\beta\xi(t)(|1001\rangle-|0110\rangle)_{u_{1}u_{_{2}}v_{1}v_{_{2}}},
\end{eqnarray}
where $\xi(t)$ is $\xi(t)=e^{-i(\omega_{m}+\Delta)t}(1-e^{igt})$. Therefore, the maximally entangled state between four mechanical resonators $u_{1}$, $u_{2}$, $v_{1}$ and $v_{2}$
\begin{eqnarray}      \label{eqH14}
|\psi\rangle_{_{2}}=\frac{1}{\sqrt{2}}(|1001\rangle-|0110\rangle)_{u_{1}u_{_{2}}v_{1}v_{_{2}}},
\end{eqnarray}
can be obtained with successful probability
\begin{eqnarray}      \label{eqH8}
P=2|\alpha\beta|^{^{2}}\sin^{2}(\frac{g_{}t}{2}).
\end{eqnarray}
At time $t=\frac{(2n_{1}+1)\pi}{g}$ \cite{ChenOE2017}, we can get the maximal probability
\begin{eqnarray}      \label{eqH9}
P_{max}=2|\alpha\beta|^{^{2}}.
\end{eqnarray}

\begin{figure}[!ht]
\begin{center}
\includegraphics[width=8.0cm,angle=0]{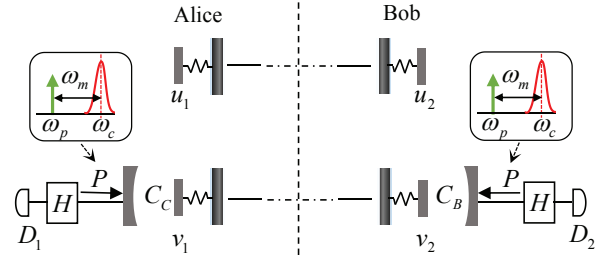}
\caption{Schematic diagram of steps 2-4 in the ECP. $u_{j}$ ($v_{j}$) (j=1,2) are 4 mechanical oscillators, among which $v_{1}$ ($v_{2}$) acts as the end mirror of the Fabry-Perot cavity $C_{C}$ ($C_{B}$). Letter $P$ means the input pump laser used for driving the anti-Stokes sideband interaction in the photon-phonon systems. The inset shows the frequency relationship between the pump field (green arrow) and the cavity mode (red line). Letter $H$ represents Hadamard gate operation of a photon in step 3, and $D_{j}$ ($j=1,2$) is the photon detectors required in step 4. All elements on the left of vertical dashed line are used by Alice, while those on the right by Bob.}\label{fig2}
\end{center}
\end{figure}

In step 2, as shown in Fig.~\ref{fig2}, two red-detuned pump pulses are input to cavities $B$ and $C$ to drive the anti-Stokes transition, respectively. In this process, an anti-Stokes photon is emitted with annihilating a phonon. In fact, it has been well realized experimentally to transform a phonon into a photon via an anti-Stokes process \cite{RiedingerNature2018}. The corresponding interaction Hamiltonian is
\begin{eqnarray}      \label{eqH}
H_{as}=G\hat{c_{j}}\hat{v_{j}}^{\dag}+ H.c.,
\end{eqnarray}
where $\hat{c}_j$ ($j=1,2$) is the photon annihilation operator of the cavity mode $C$ if $j=1$ or $B$ if $j=2$, and $\hat{v}_j$ ($j=1,2$) is the phonon annihilation operator of the resonator mode $v_j$. $G$ is the coupling strength between the $j$th mechanical resonator and the driven cavity after linearizing treatment. With this Hamiltonian, the mechanical resonator state, Eq. (\ref{eqH14}), can then be mapped onto a phonon-photon state
\begin{eqnarray}      \label{eqH}
|\psi\rangle_{_{3}}=\frac{1}{\sqrt{2}}(|10\rangle_{u_{1}u_{2}}|01\rangle_{CB}-|01\rangle_{u_{1}u_{2}}|10\rangle_{CB}).
\end{eqnarray}

In step 3, as shown in Fig.~\ref{fig2}, Alice and Bob perform a Hadamard gate operation on photons $C$ and $B$, respectively. The state of the composite system becomes
\begin{eqnarray}      \label{eqH}
|\psi\rangle_{_{4}}&=&\frac{1}{2\sqrt{2}}\Big[(|10\rangle_{u_{1}u_{2}}-|01\rangle_{u_{1}u_{2}})\nonumber\\
&&\otimes(|00\rangle_{CB}-|11\rangle_{CB})\nonumber\\
&&+(|10\rangle_{u_{1}u_{2}}+|01\rangle_{u_{1}u_{2}})\nonumber\\
&&\otimes(|10\rangle_{CB}-|01\rangle_{CB})\Big].
\end{eqnarray}

In final step 4, Alice and Bob use photondetectors to detect the photons and share the detection results. Up to now, the ECP is accomplished. What they obtain is the maximally entangled phonon state
\begin{eqnarray}      \label{eqH}
|\psi\rangle_{_{5}}=\frac{1}{\sqrt{2}}(|10\rangle_{u_{1}u_{2}}+|10\rangle_{u_{1}u_{2}}),
\end{eqnarray}
if their photon counting results are different, or
\begin{eqnarray}      \label{eqH}
|\psi\rangle_{_{6}}=\frac{1}{\sqrt{2}}(|10\rangle_{u_{1}u_{2}}-|01\rangle_{u_{1}u_{2}}),
\end{eqnarray}
if the photon counting results are same. The photon state $|\psi\rangle_{6}$ can be transferred to $|\psi\rangle_{5}$ via a $\pi$-phase operation about anyone of phonons. All the steps of entanglement concentration shown above are listed in Table~\ref{tab}.

\begin{table}
\centering
\caption{The core steps of the ECP}
\begin{tabular}{l c p{0.3cm} c r}
\hline\hline
&Step&& Process&\\
\hline
&1&& postselect maximally entangled 4-phonon state&\\
&2&& transfer phonons to photons via anti-Stokes&\\
&3&& make Hadamard gate operations on photons&\\
&4&& select Bell (GHZ) state&\\
\hline\hline\label{tab}
\end{tabular}
\end{table}

\section{Entanglement concentration for less-entangled GHZ state of phonons}\label{sec3}
\begin{figure}[!ht]
\begin{center}
\includegraphics[width=8.0cm,angle=0]{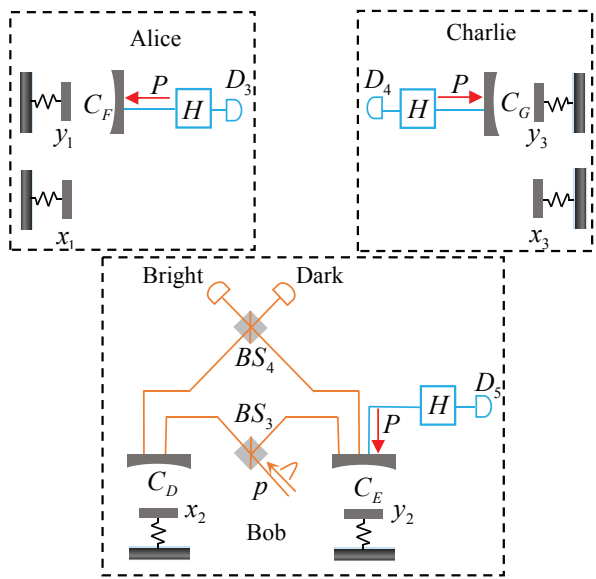}
\caption{Schematic diagram of the ECP for phonon GHZ states. The ECP is divided into three parts (see dashed-line long squares) held by Alice, Bob, and Charlie, respectively. The function of Alice' or Charlie' part is the same as that of Alice or Bob in Fig. 2, and the function of Bob' part is the same as that of Bob in Fig.~\ref{fig1} }\label{fig3}
\end{center}
\end{figure}

In this section we extend the above investigation about the two-resonator case over multi-resonator one, and consider how a less-entangled GHZ state of phonons is concentrated for entanglement. Figure~\ref{fig3} is the schematic diagram of the scheme, where the mechanical resonator $x_j$ ($y_j$) belongs to Alice if $j=1$, Bob if $j=2$ and Charlie if $j=3$. Initially, Alice, Bob, and Charlie share two pairs of less-entangled tripartite GHZ states
\begin{eqnarray}      \label{eqH13}
|\varphi\rangle_{x_{1}x_{2}x_{3}}&=&\alpha|000\rangle_{x_{1}x_{2}x_{3}}+\beta|111\rangle_{x_{1}x_{2}x_{3}}\nonumber\\
|\varphi\rangle_{y_{1}y_{2}y_{3}}&=&\alpha|000\rangle_{y_{1}y_{2}y_{3}}+\beta|111\rangle_{y_{1}y_{2}y_{3}},
\end{eqnarray}
with $|\alpha|^{2}+|\beta|^{2}=1$. Firstly, Bob combines the optomechanical cross-Kerr interaction with the Mach-Zehnder interferometer to realize the maximal entanglement between the six mechanical resonators by detecting an output photon at the dark port. A photon is then sent to the interferometer of $BS_3$, and its state after $BS_3$ becomes $|\varphi\rangle_{in}=\frac{1}{\sqrt{2}}(|10\rangle_{DE}+|01\rangle_{DE})$. When a photon enters cavity $D$ or $E$, the composite state of the system will evolve to
\begin{widetext}
\begin{eqnarray}      \label{eqH}
|\varphi(t)\rangle&=&|\varphi\rangle_{i}\otimes|\varphi\rangle_{x_{1}x_{2}x_{3}}\otimes|\varphi\rangle_{y_{1}y_{2}y_{3}}\nonumber\\
\longrightarrow&&\frac{1}{\sqrt{2}}\Big[|10\rangle_{DE}\otimes(\alpha e^{-i\theta_{10}}|000\rangle_{x_{1}x_{2}x_{3}}+\beta e^{-i\theta_{11}}|111\rangle_{x_{1}x_{2}x_{3}})\otimes(\alpha|000\rangle_{y_{1}y_{2}y_{3}}+\beta e^{^{-i\theta_{01}}}|111\rangle_{y_{1}y_{2}y_{3}})\nonumber\\
&&+|01\rangle_{DE}\otimes(\alpha|000\rangle_{x_{1}x_{2}x_{3}}+\beta e^{-i\theta_{01}}|111\rangle_{x_{1}x_{2}x_{3}})\otimes(\alpha e^{-i\theta_{10}}|000\rangle_{y_{1}y_{2}y_{3}}+\beta e^{^{-i\theta_{11}}}|111\rangle_{y_{1}y_{2}y_{3}})\Big].
\end{eqnarray}
\end{widetext}
If an output photon is detected at the dark port, the single-photon state $|\varphi\rangle_{f}=\frac{1}{\sqrt{2}}(|10\rangle_{DE}-|01\rangle_{DE})$ is postselected. The final state of the six mechanical resonators is
\begin{eqnarray}      \label{eqH}
|\varphi\rangle_{1}&=&\frac{1}{2}\alpha\beta e^{-i(\omega_{m}+\Delta)t}(1-e^{igt})\nonumber\\
&&\times(|000111\rangle_{x_{1}x_{2}x_{3}y_{1}y_{2}y_{3}}\nonumber\\
&&-|111000\rangle_{x_{1}x_{2}x_{3}y_{1}y_{2}y_{3}}).
\end{eqnarray}
The maximally entangled state can be obtained with a successful probability of postselection, Eq.(\ref{eqH8}), and reads
\begin{eqnarray}      \label{eqH}
|\varphi\rangle_{2}&=&\frac{1}{\sqrt{2}}(|000111\rangle_{x_{1}x_{2}x_{3}y_{1}y_{2}y_{3}}\nonumber\\
&&-|111000\rangle_{x_{1}x_{2}x_{3}y_{1}y_{2}y_{3}}).
\end{eqnarray}
At time  $t=\frac{2(n_{1}+1)\pi}{g}$, the successful probability becomes maximal as shown by Eq.(\ref{eqH9}).

Secondly, each of these three cavities $E$, $F$ and $G$ are pumped with an
anti-Stokes light respectively to project the mechanical
state into the optical cavity state
\begin{eqnarray}      \label{eqH}
|\varphi\rangle_{3}&=&\frac{1}{\sqrt{2}}(|000\rangle_{x_{1}x_{2}x_{3}}|111\rangle_{EFG}\nonumber\\
&&-|111\rangle_{x_{1}x_{2}x_{3}}|000\rangle_{EFG}).
\end{eqnarray}
Here $|i\rangle_{c_j}$ ($i=0,1$ and $j=2,3,4$ for $c_2=E$, $c_3=F$, and $c_4=G$) is the projected Fock state of $i$ photon in cavity $c_{j}$. Then all the users make a Hadamard gate operation on their photons and get
\begin{eqnarray}      \label{eqH}
|\varphi\rangle_{4}&=&\frac{1}{4}\Big[|000\rangle_{x_{1}x_{2}x_{3}}(|0\rangle_{E}-|1\rangle_{E})\nonumber\\
&&\times(|0\rangle_{F}-|1\rangle_{F})(|0\rangle_{G}-|1\rangle_{G})\nonumber\\
&&-|111\rangle_{x_{1}x_{2}x_{3}}(|0\rangle_{E}+|1\rangle_{E})\nonumber\\
&&\times(|0\rangle_{F}+|1\rangle_{F})(|0\rangle_{G}+|1\rangle_{G})\Big].
\end{eqnarray}
When the photon counting is an odd number of $|1\rangle$ after detection, the state of the mechanical resonators collapses to
\begin{eqnarray}      \label{eqH21}
|\varphi\rangle_{5}&=&\frac{1}{\sqrt{2}}(|000\rangle_{x_{1}x_{2}x_{3}}+|111\rangle_{x_{1}x_{2}x_{3}}).
\end{eqnarray}
On the contrary, if the photon counting is an even number of $|1\rangle$, the state becomes
\begin{eqnarray}      \label{eqH22}
|\varphi\rangle_{6}&=&\frac{1}{\sqrt{2}}(|000\rangle_{x_{1}x_{2}x_{3}}-|111\rangle_{x_{1}x_{2}x_{3}}).
\end{eqnarray}
The photon state Eq.(\ref{eqH22}) can be transferred to state Eq.(\ref{eqH21}) and vice versa via a $\pi$-phase operation performed by anyone of these three users about phonons. By now, we have accomplished the entanglement concentration for the less-entangled GHE state Eq. (\ref{eqH13}).

\begin{figure}[!ht]
\begin{center}
\includegraphics[width=7cm,angle=0]{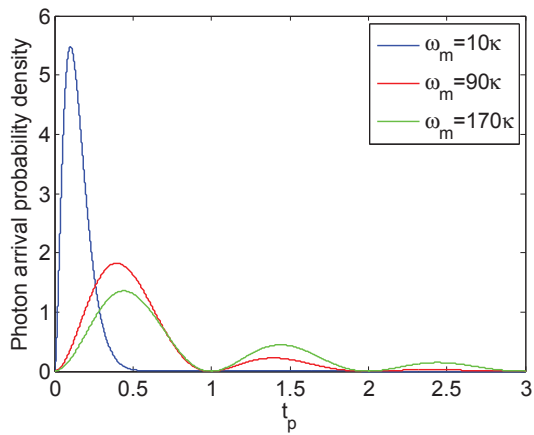}
\caption{ Photon arrival probability density vs arrival time for a successful postselection in the sideband-resolved regime $\omega_{m1}=30\kappa$ (green), $\omega_{m}=90\kappa$ (red)
and $\omega_{m}=150\kappa$ (blue). Here $\omega_{m}=2\pi$ GHz, $g=3.33\times10^{-2}\omega_{m}$.
}\label{figarrivalrate}
\end{center}
\end{figure}

Before ending this section, we make two remarks to the experimental feasibility about our protocol. The first is that in our protocol, the optomechanical interaction plays an essential role, because the quantum nondemolition measurement and the gate operations of phonons require that the mechanical resonators must be operated indirectly, while this could easily be achieved via optomechanical interaction. Recent investigations have confirmed the possibility of the above phonon operation by applying the cross-Kerr effect between photons and phonons via optomechanical interaction \cite{ChenOE2017,YinOL2018,Liaoarxiv2018}. Second, it is stated that photon-postselection performed at the dark-port of a bean-splitter in step 1 is an important operation of the ECP for the multiuser GHZ state. We can use the quantity, the photon arrival probability density (PAPD), to estimate the probability of a successful postselection. The PAPD can be calculated based on the formula \cite{PepperPRL2012}
\begin{eqnarray}        \label{eqHrate}
\frac{2|\alpha\beta|^{^{2}}\sin^{2}(\frac{g_{}t}{2})\kappa exp(-\kappa t)}{P_{tot}},
\end{eqnarray}
where $t$ is the time after the photon releasing from an optical cavity, $\kappa$ the decay of the cavity, $\kappa exp(-\kappa t)$ the probability density of a photon, and $2|\alpha\beta|^{^{2}}\sin^{2}(\frac{g_{}t}{2})$  the probability of a successful postselection at $t$. $P_{tot}$ is the overall single photon probability for creating the state $|\psi_{2}\rangle$ shown by Eq.(\ref{eqH14}) and is described by
\begin{eqnarray}        \label{eqH46}
P_{tot}=2|\alpha\beta|^{^{2}}\kappa\!\!\int^{\infty}_{0}\!\!\!\!\sin^{2}(\frac{g_{}t}{2})exp(-\kappa t)dt=|\alpha\beta|^{^{2}}\frac{g^{2}}{g^{2}+\kappa^{2}}.\nonumber\\
\end{eqnarray}
We plot the PAPD versus time $t_{p}$ ($=\frac{gt}{2\pi}$) in Fig.~\ref{figarrivalrate}.
It is shown that $\omega_{m}\geq90\kappa$ should be satisfied to obtain the observable oscillations of the arrival rate in an optomechanical cavity. The scheme should work in the
resolved-sideband regime if the noises from the dark count rate of the detector are taken into consideration. For instance, as we choose $\omega_{m}=2\pi$ GHz which can be realized by a suspended bulk acoustic resonator \cite{RiedingerNature2018}, $g=3.33\times10^{-2}\omega_{m}$ and $\kappa=1/90\omega_{m}$, the probability of the successful postselection, Eq.(\ref{eqH46}), is approximately $0.8997|\alpha\beta|^{^{2}}$.  The window for detectors receiving photons is approximately $1/\kappa$, thus the dark count rate should be less than $0.8997|\alpha\beta|^{^{2}}\kappa$. The current best silicon avalanche photodiodes have a dark count rate of $\sim2$Hz, which means that $|\alpha\beta|^{^{2}}\geq3.184\times10^{-8}$ is need to be satisfied for the optomechanical device with $\omega_{m}=90\kappa$.

\section{summary} \label{sec4}
In summary, we proposed a protocol for nonlocal-phonon entanglement concentration both for the Bell state and the GHZ state. We use the optomechanical cross-Kerr interaction and the Mach-Zehnder interferometer to postselect two phonon pairs with maximally entangled states by detecting the photon output at the dark port of the interferometer. After transforming phonons to photons via another optomichanical interaction and making the Bell-state analysis, we derive the maximally entangled phonon states shared by nonlocal users, such as the Bell state and the GHZ state, by making the Bell-state analysis. Entanglement is a basic resource for various QIPs and our work is useful in realizing those phonon-based QIPs.

\section*{ACKNOWLEDGMENT}
This work was supported by National Natural Science Foundation of China under Grants No. 11654003 and No. 11174040; Fundamental Research Funds for the Central Universities (2017TZ01); The Interdiscipline Research Funds of Beijing Normal University.

\bigskip \appendix
\section*{appendix: ENTANGLEMENT PREPARATION FOR TWO REMOTE MECHANICAL RESONATORS}
\setcounter{equation}{0}  



\renewcommand{\theequation}{ A\arabic{equation}}

\begin{figure}[!ht]
\begin{center}
\includegraphics[width=7cm,angle=0]{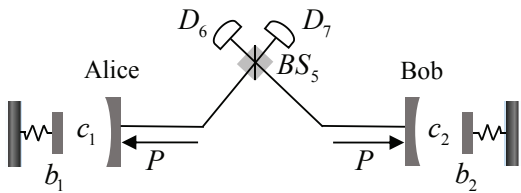}
\caption{Schematic diagram of entanglement creation for nonlocal phonons. Two pump pulses described by letter $P$ are input to realize a squeezing interaction between the mechanical mode $b_{i}$ and the optical mode $c_{i}$ ($i=1,2$) in the optomechanical systems shared by Alice and Bob, respectively. }\label{fig5}
\end{center}
\end{figure}

Ref.\cite{RiedingerNature2018} as reported experimental generation of entanglement between nonlocal phonons with two remote optomechanical systems. We persent in this appendix the theoretical model about the Bell state generation for nonlocal mechanical resonators. Assuming two separate optomechanical systems owned by Alice and Bob are set symmetrically (see Fig.~\ref{fig5}). Each of them consists of an optical cavity and a mechanical resonator, and is driven by a pump pulse near the blue sideband of the cavity to stimulate the Stokes process. After the standard linearization procedure, the corresponding interaction Hamiltonian can be written as
\begin{eqnarray}
\hat{H}_{s}= G\hat{c}_{i}^{\dag}\hat{b}_{i}^{\dag}+ H.c.,
\end{eqnarray}
where $\hat{c}_{i}^{\dag}$ and $\hat{b}_{i}^{\dag}$ ($i=1,2$) represent, respectively, the creation operators for the $i$-th cavity photon and the $i$-th mechanical resonator, $G=g_{0}\sqrt{n}$ is the effective linear coupling strength which can be varied by changing the intracavity photon number $n$ and the single-photon coupling $g_{{0}}$. With the low energy pump pulse \cite{RiedingerNature2018}, the composite state of these two optomechanical systems is described by
\begin{eqnarray}
|\phi\rangle_{1}\otimes|\phi\rangle_{_{2}}\!&\!=\!&\!(|0\rangle_{c_{1}}|0\rangle_{b_{1}}\!+\!\sqrt{p_{p}}\hat{c}^{\dag}_{1}\hat{b}^{\dag}_{1}|0\rangle_{c_{1}}|0\rangle_{b_{1}}) \nonumber\\
\!&\!\!&\!\otimes(|0\rangle_{c_{2}}|0\rangle_{b_{2}}\!+\!\sqrt{p_{p}}\hat{c}^{\dag}_{2}\hat{b}^{\dag}_{2}|0\rangle_{c_{2}}|0\rangle_{b_{2}}),
\end{eqnarray}
where $|j\rangle_{c_i}$ ($|j\rangle_{b_i}$) ($i=1,2$ or $j=0,1$) represent $j$ photons in the cavity mode $c_i$ ($j$ phonons in the resonator mode $b_i$). The optomechanical system with subscript $i$ is hold by Alice if $i=1$ or by Bob if $i=2$.  The scattered Stokes photons from two optomechanical systems interfere at beam splitter $BS_5$ with relation $\hat{c}_{\pm}=(c_{1}\pm c_{2})/\sqrt{2}$. After the $BS_5$, the composite
state will evolve to
\begin{eqnarray}
|\phi\rangle_{1}\!\otimes\!|\phi\rangle_{_{2}}\!\!&\!=\!&\!\!\sqrt{p_{p}}\Big[\!\frac{1}{\sqrt{2}}\hat{c}^{\dag}_{+}(\hat{b}^{\dag}_{1}\!+\!\hat{b}^{\dag}_{2})
\!+\!\frac{1}{\sqrt{2}}\hat{c}^{\dag}_{-}(\hat{b}^{\dag}_{1}\!-\!\hat{b}^{\dag}_{2})\!+\!1\!\Big]\!|0\rangle,\nonumber\\
\end{eqnarray}
where $|0\rangle$ means the vacuum state $|0\rangle=|00\rangle_{c_{1}c_{2}}|00\rangle_{b_{1}b_{2}}$ and the term $\hat{c}^{\dag}_{2}\hat{b}^{\dag}_{2}\hat{c}^{\dag}_{1}\hat{b}^{\dag}_{1}|0\rangle$ has been neglected due to the low scattering probability. When there is a click in the photondetector $D_6$ or $D_7$, the projected state of the mechanical resonator 1 and 2 is
\begin{eqnarray}
|\psi\rangle^{\pm}_{b_1b_2}=\frac{1}{\sqrt{2}}(|1\rangle_{b_{1}}|0\rangle_{b_{2}}\pm|0\rangle_{b_{1}}|1\rangle_{b_{2}}).
\end{eqnarray}
Above Bell states are usually used for quantum communications.

\end{document}